# Structure, microstructure and hydrogen storage properties of melt-spun $V_{55}Ti_{21}Cr_{17}Fe_7$ and $V_{55}Ti_{21}Mn_{17}Fe_7$

A.Ioannidou[1,2], S.S. Makridis[1,2], M. Gjoka[2], E.I. Gkanas[1,2] A.K. Stubos[2], N. Lupu[3], D. Niarchos[2]

[1]Department of Mechanical Engineering, University of Western Macedonia, Kozani, 50100, GREECE.

[2]National Center of Scientific Research 'Demokritos', Agia Paraskevi, Athens, 15310, GREECE.

National Center of Scientific Research 'Demokritos', Agia Paraskevi, Athens, 15310, GREECE.

[3]National Institute of Research and Development for Technical Physics, Iasi, Romania

**Abstract:**

The hydrogen sorption performance of $V_{55}Ti_{21}Cr_{17}Fe_7$ and $V_{55}Ti_{21}Mn_{17}Fe_7$ alloys and their ribbons were evaluated by pressure-composition-temperature tests. Their hydrogen absorption kinetic properties were studied through hydrogen absorption curves. The crystallographic structures and microstructure of these alloys and ribbons before hydrogen absorption and after hydrogen desorption PCT tests were identified by X-ray diffraction (XRD) and Scanning electron microscopy (*SEM*) analysis, respectively. Hydrogen storage characteristics of such materials were investigated by volumetric method using Sievert's type apparatus and gravimetric method with suspension balance.

## 1. Introduction

Energy systems based on hydrogen are interesting candidates to replace fossil fuels for the future energy global needs and assist in mitigating the associated environmental problems [1]. The absorption of hydrogen in solid materials and particularly in metal hydrides seems to be a safe and promising technique, due to the high volumetric energy density and the required relatively low pressures [2]. Hydrogen storage alloys can be classified as $AB_5$–type (e.g. $LaNi_5$), $AB_2$-type (e.g. Ti-Zr alloys), $A_2B$-type (e.g. $Sb_2Ti$, $Sn_2Co$) and AB-type (e.g. Ti-Fe alloys) [3-5], but their storage capacity is relatively low (less than 2 wt%) [6]. Mg-based alloys have higher capacities (of the order of 6.5%), but the absorption/desorption processes take place at a high temperature range (300-400$^0$C) [7]. It is therefore imperative that efforts be made to develop new hydrides with larger hydrogen storage capacities at room temperature [8].

Recently, Vanadium-based solid solution alloys with B.C.C crystal structure have been identified as potential materials for hydrogen storage, because they present remarkable storage capacities of 3.8% wt [9-12]. V-Ti-Cr alloys are the most known B.C.C solid solution alloys which are used for hydrogen storage applications. However, such materials have some disadvantages such as high cost, difficult activation process, slow kinetics and large hysteresis between absorption and desorption [13-15].

Researchers over the past few years tried to develop techniques to increase the usable hydrogen in the alloys and to raise the desorption plateau pressure improving the desorption capacity by adding Fe and Ni or Fe and Mn respectively inside V-Ti-Cr alloys [16], or by partial substitution of V by Fe in $Ti_{10}Cr_{18}Mn_{27}V$ alloys [17]. Further, it was demonstrated that the hydrogen storage properties of V(30 at%)-Ti(15-55 at%)-Cr(7-43 at%)-Fe(2-18 at%) depend on the Ti/(Cr+Fe) ratio [18] and the Laves phase depends on the x content on $V_x$-$(Ti-Cr-Fe)_{100-x}$ alloys [19]. Further, it was reported that rapid solidification is also an effective method to improve the kinetics of such hydrogen storage alloys. It was found that solidification process refined a dendritic microstructure and the absorption capacity was increased [20, 21].

In the present work, the hydrogen desorption performance of $V_{55}Ti_{21}Cr_{17}Fe_7$ and $V_{55}Ti_{21}Mn_{17}Fe_7$ alloys and their ribbons was investigated. It should be noted that to the best knowledge of the authors, there are no references concerning the hydrogen desorption performance of such materials, so this study offers new data to the field of hydrogen storage in B.C.C solid solutions. XRD and SEM analysis was also performed in order to identify the microstructural characteristics of these materials.

2. Experimental procedure

The samples with nominal composition $V_{55}Ti_{21}X_{17}Fe_7$ (where X= Cr, Mn) were prepared from pure metals giving rise to ribbons in different wheel speeds. X-ray analysis with Cu-K$\alpha$ radiation was used to identify the structure of the samples while SEM with energy dispersive spectroscopy (EDS) was used for analyzing the microstructure characteristics and the average chemical composition of the alloys. An activation procedure was performed before hydrogenation measurements. Pressure–composition (P–C) isotherms were measured by the Sievert's method with the use of a magnetic suspension balance at 298K under 4 MPa.

3. Results and Discussion

**X-ray results**

XRD patterns of $V_{55}Ti_{21}X_{17}Fe_7$ (where X=Mn, Cr) are shown in Fig. 1 and Fig.2. Rietveld analysis showed that the compound with nominal composition $V_{55}Ti_{21}Mn_{17}Fe_7$ has a single BCC cubic phase structure while in the Cr substituted alloy a secondary cubic phase appeared. In both samples a preferred orientation arises on 002 hkl due to the stronger tendency of the crystallites to be oriented in more than one way in the powder. The lattice parameters of the major phase remained the same after substituting Cr with Mn which was expected according to [22]. The results of refinement are shown in Table 1. A sample with nominal composition $V_{55}Ti_{21}Cr_{17}Fe_7$ was examined using SEM. EDS analysis showed that the alloy follows the nominal composition with the cubic matrix phase while a second phase appears which is a V rich-phase shown in the area 1.13 in

Fig. 3 (b). These results are in good agreement with the XRD results and previous studies published by other groups for V rich alloys [19].

## Hydrogen desorption properties

Hydrogen desorption measurements were performed for the ribbons $V_{55}Ti_{21}Cr_{17}Fe_7$ and $V_{55}Ti_{21}Mn_{17}Fe_7$ using a volumetric method with a Sievert's type apparatus. The results in Fig. 4 show that these materials can desorb almost 3 wt% of hydrogen at room temperature under 3.5 MPa. Comparing the measurements between the Cr substituted alloys, the sample made at 20 m/s absorbed more hydrogen than the one made at 25 m/s. On the contrary, for the samples made at the same speed of 25 m/s the Mn-substituted alloy desorbed more hydrogen at the same temperature. In order to confirm this measurement a gravimetric method was also employed using a magnetic suspension balance. Fig. 5 shows the gravimetric test of $V_{55}Ti_{21}Mn_{17}Fe_7$ where the desorbed hydrogen is almost the same as measured with the volumetric method.

**Conclusions**

The hydrogen storage properties of $V_{55}Ti_{21}Mn_{17}Fe_7$ and $V_{55}Ti_{21}Cr_{17}Fe_7$ alloys in ribbon form were investigated for the first time. Both samples exhibit B.C.C cubic phase with a secondary phase for the Cr substituted phase. Hydrogenation procedures followed by two different measurement methods showed that the compound with stoichiometry $V_{55}Ti_{21}Mn_{17}Fe_7$ made at 25 m/s has the highest desorption capacity.

**Acknowledgements**

This work has been supported from the Greek Ministry of Scientific Research and Technology, Greece-Romania bilateral project, code: **11ROM 1_3_ET 30**

**References**

[1] Princini G, Agresti F, Maddalena A, Lo Russo S. The problem of solid state hydrogen storage. *Energy* 2009; **34**:2087-2091.

[2] Gambini M, Manno M, Vellini M. Numerical study and performance assestment of metal hydride – based hydrogen storage systems. *International Journal of Hydrogen Energy* 2008; **33**:6178-6187.

[3] Larcher D, Beaulieu LY, Mao O, George AE, Dahn JR. Study of the Reaction of Lithium with Isostructural $A_2B$ and Various $Al_xB$ Alloys. *Journal of Electrochemistry Society* 2000; **147**(5):1703-1708.

[4] Vincent Bérubé, Gregg Radtke, Mildred Dresselhaus, Gang Chen. Size effects on the hydrogen storage properties of nanostructured metal hydrides: A review. *International Journal of Energy Research* 2007; **31**(6-7): 637–663.

[5] Lee H-H, Lee K-Y, Lee J-Y. The hydrogenation characteristics of Ti-Zr-V-Mn-Ni C14 type Laves phase alloys for metal hydride electrodes. *Journal of Alloys and Compounds* 1997; **253–254**:601–604.

[6] Singh BK, Singh AK, Srivastava ON. Improved hydrogen sorption characteristics in $FeTi_{1+x}Mm$ material. *International Journal of Hydrogen Energy* 1996; **21**(2): 111–117.

[7] Hakan Gasan, Osman N. Celik, Nedret Aydinbeyli, Yasar M. YamaN. Effect of V, Nb, Ti and graphite additions on the hydrogen desorption temperature of magnesium hydride. *International Journal of Hydrogen Energy* 2012; **37**:1912-1918.

[8] Okada M, Kuriiwa T, Tamura T,Takamura H, Kamegawa A. Ti–V–Cr b.c.c. alloys with high protium content. *Journal of Alloys and Compounds* 2002; **330–332**: 511–516.

[9] Huot J, Enoki H, Akiba E. Synthesis, phase transformation, and hydrogen storage properties of ball-milled $TiV_{0.9}Mn_{1.1}$. *Journal of Alloys and Compounds* 2008; **453**:203-209.

[10] Dehouche Z, Savard M, Laurencelle F, Goyette J. Ti–V–Mn based alloys for hydrogen compression system. *Journal of Alloys and Compounds* 2005; **400**:276–280.

[11]Taizhong H, Zhu Wu, Baojia X , Naixin Xu. Effect of stoichiometry on hydrogen storage performance of Ti–Cr–V–Fe based alloys. *Intermetallic* 2005; **13**(10):1075–8.

[12] Lynch JF, Libowitz GG, Maeland AJ. Use of vanadium-based solid solution alloys in metal hydride heat pumps. *Journal of Less Common Metals* 1987; **131**:275.

[13] Itoh H, Arashima H, Kubo K, Kabutomori T. The influence of microstructure on hydrogen absorption properties of Ti–Cr–V alloy. *Journal of Alloys and Compounds* 2002; **330–332**:287–91.

[14] Esayed AY, Northwood DO. The effect of hydride formation and decomposition cycles on hysteresis in $V_xNb_{1-x}$-H systems. *International Journal of Hydrogen Energy* 1992; **17**: 211.

[15] Mouri T, Iba H. Hydrogen-absorbing alloys with a large capacity for a new energy carrier. *Materials Science and Engineering: A* 2002; **329–331**:346–350.

[16] Liang Hao, Chen Yungui, Yan Yigang, Wu Chaoling, Tao Mingda. Influence of Ni or Mn on hydrogen absorption–desorption performance of V–Ti–Cr–Fe alloys. *Materials Science and Engineering A* 2007; **459**:204–208.

[17] Yu XB, Yang ZX, Feng SL, Wu Z, Xu NX. Influence of Fe addition on hydrogen storage characteristics of Ti–V-based alloy. International Journal of Hydrogen Energy 2006; **31**:1176–1181.

[18] Yigang Yan, Yungui Chen, Hao Liang, Chaoling Wu, Mingda Tao. Hydrogen storage properties of V30–Ti–Cr–Fe alloys. *Journal of Alloys and Compounds* 2007; **427**:110–114.

[19] Yigang Yan, Yungui Chen, Hao Liang, Xiaoxiao Zhou, Chaoling Wu, Mingda Tao, Lijuan Pang. Hydrogen storage properties of V–Ti–Cr–Fe alloys. *Journal of Alloys and Compounds* 2008; **454**: 427–434.

[20] Yunfeng Zhu, Yanfang Liu, Feng Hua, Liquan Li. Effect of rapid solidification on the structural and electrochemical properties of the Ti–V-based hydrogen storage electrode alloy. *Journal of Alloys and Compounds* 2008; **463**:528-532.

[21] P. Pei, X.P. Song, J. Liu, G.L. Chen, X.B. Qin, B.Y. Wang. The effect of rapid solidification on the microstructure and hydrogen storage properties of $V_{35}Ti_{25}Cr_{40}$ hydrogen storage alloy. *International Journal of Hydrogen Energy* 2009; **34**:8094-8100.

[22] Jeong-Hyun Yoo, Gunchoo Shim, Choong-Nyeon Park, Won-Baek Kim, Sung-Wook Cho. Influence of Mn or Mn plus Fe on the hydrogen storage properties of the Ti-Cr-V alloy. *International Journal of Hydrogen Energy* 2009; **34**:9116-9121.

Table 1. Rietveld Analysis and crystallographic characteristics of the phases

| Alloy | Space group | Lattice | | | | | | | |
|---|---|---|---|---|---|---|---|---|---|
| V$_{55}$Ti$_{21}$Cr$_{17}$Fe$_7$ | Pm3m | a=b=c=3.021 | 2.52 | 93.57 | 12.71 | 17.36 | 13.39 | 0.5646 on hkl 002 |
| | Fm3m | a=b=c=4.9121 | 5.68 | 6.43 | | | | |
| V$_{55}$Ti$_{21}$Mn$_{17}$Fe$_7$ | Pm3m | a=b=c=3.012 | 5.62 | 100 | 14.17 | 18.01 | 12.45 | 0.6076 on hkl 002 |

Table 2: SEM results of alloys (Atomic %)

| | V$_{55}$Ti$_{21}$Cr$_{17}$Fe$_7$ | | V$_{55}$Ti$_{21}$Mn$_{17}$Fe$_7$ |
|---|---|---|---|
| Element | Photo (a) | Photo (b) | Photo (c) |
| Ti | 20.72 | 20.40 | 21.12 |
| V | 52.55 | 73.22 | 54.50 |
| Cr , Mn | 17.63 | 5.12 | 17.2 |
| Fe | 9.10 | 1.26 | 7.18 |

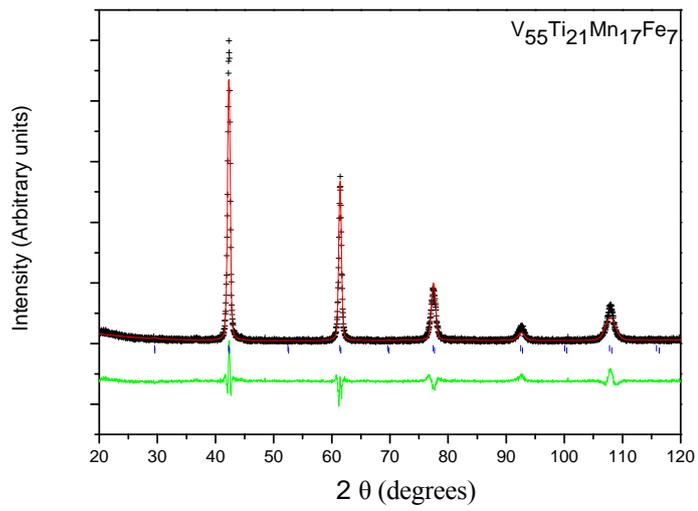

**Figure 1. XRD pattern for $V_{55}Ti_{21}Mn_{17}Fe_7$ alloy.**

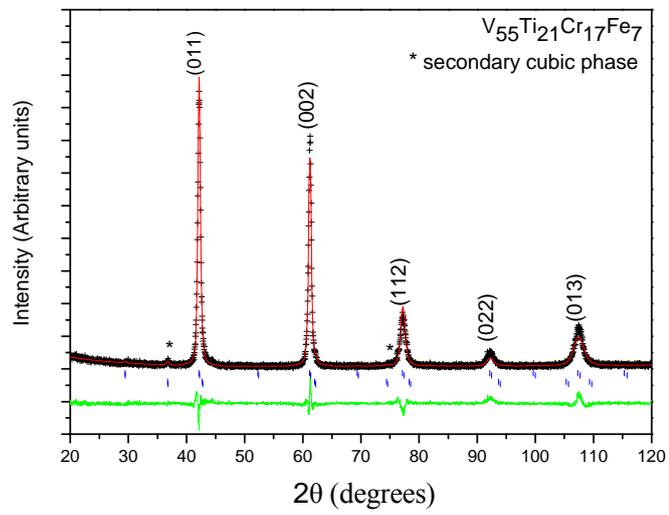

**Figure 2. XRD pattern for $V_{55}Ti_{21}Cr_{17}Fe_7$ alloy.**

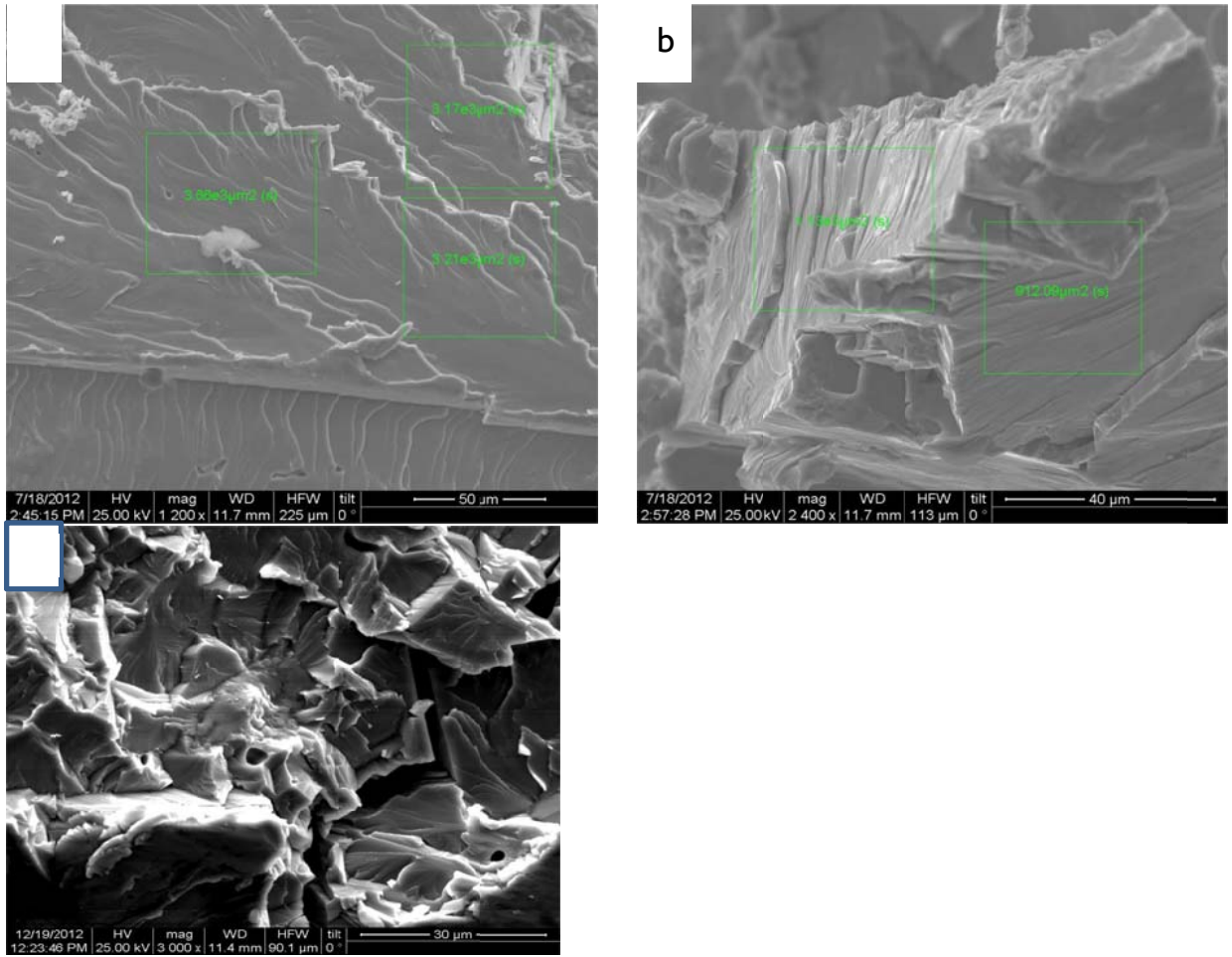

**Figure 3. SEM micrographs of alloys $V_{55}Ti_{21}Cr_{17}Fe_7$ ( a and b) and $V_{55}Ti_{21}Mn_{17}Fe_7$. (c) alloys.**

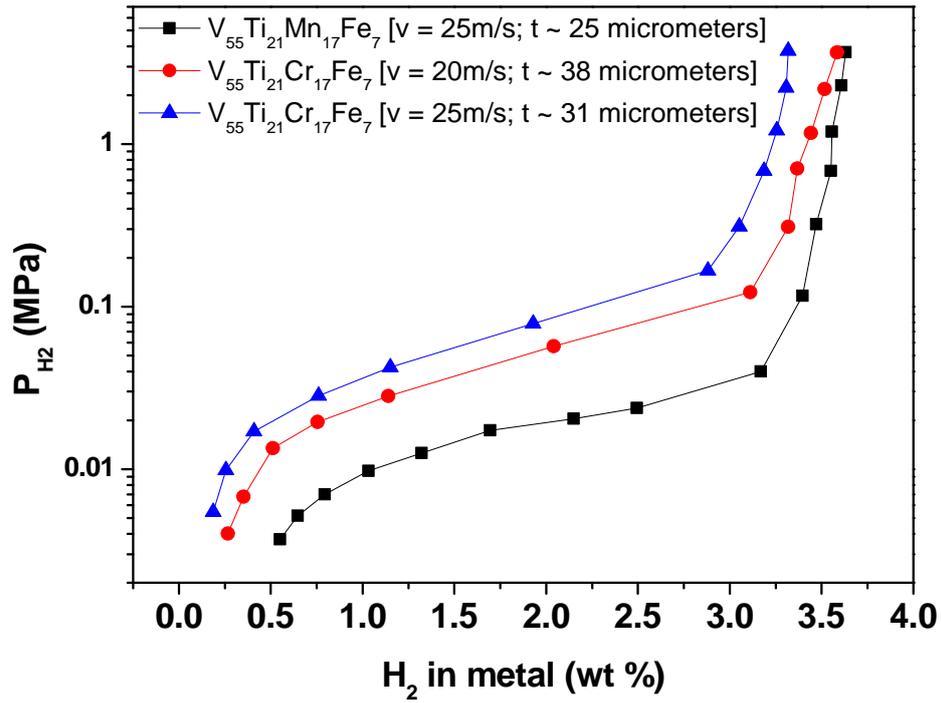

**Figure 4.** Hydrogen desorption isotherm measurements of $V_{55}Ti_{21}Mn_{17}Fe_7$ (black line) and $V_{55}Ti_{21}Cr_{17}Fe_7$ (blue and red lines) ribbons using volumetric system

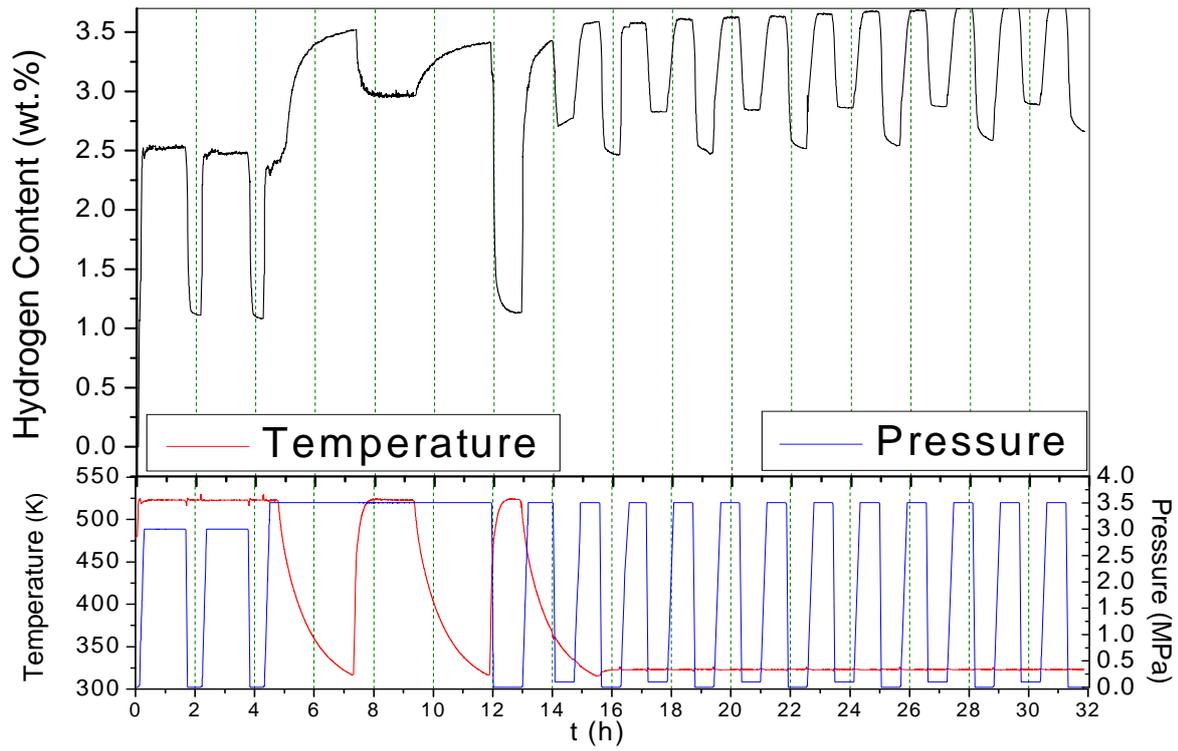

**Figure 5 :** Activation procedure and desoprtion measurements of Mn-substituted alloys using gravimetric system